\documentclass[12pt,epsfig]{article}
\usepackage{epsfig}
\begin{document}
\setlength{\baselineskip}{0.30in}
\newcommand{\nc}{\newcommand}
\newcommand{\beq}{\begin{equation}}
\newcommand{\eeq}{\end{equation}}
\newcommand{\be}{\begin{eqnarray}}
\newcommand{\ee}{\end{eqnarray}}
\newcommand{\num}{\nu_\mu}
\newcommand{\nue}{\nu_e}
\newcommand{\nut}{\nu_\tau}
\newcommand{\nus}{\nu_s}
\newcommand{\mnus}{m_{\nu_s}}
\newcommand{\taus}{\tau_{\nu_s}}
\newcommand{\nnt}{n_{\nu_\tau}}
\newcommand{\rnt}{\rho_{\nu_\tau}}
\newcommand{\mnt}{m_{\nu_\tau}}
\newcommand{\tnt}{\tau_{\nu_\tau}}
\newcommand{\bi}{\bibitem}
\newcommand{\rar}{\rightarrow}
\newcommand{\lar}{\leftarrow}
\newcommand{\lrar}{\leftrightarrow}
\newcommand{\dm}{\delta m^2}
\newcommand{\so}{\, \mbox{sin}\Omega}
\newcommand{\co}{\, \mbox{cos}\Omega}
\newcommand{\sotil}{\, \mbox{sin}\tilde\Omega}
\newcommand{\cotil}{\, \mbox{cos}\tilde\Omega}
\makeatletter
\def\alt{\mathrel{\mathpalette\vereq<}}
\def\vereq#1#2{\lower3pt\vbox{\baselineskip1.5pt \lineskip1.5pt
\ialign{$\m@th#1\hfill##\hfil$\crcr#2\crcr\sim\crcr}}}
\def\agt{\mathrel{\mathpalette\vereq>}}

\newcommand{\eq}{{\rm eq}}
\newcommand{\tot}{{\rm tot}}
\newcommand{\M}{{\rm M}}
\newcommand{\coll}{{\rm coll}}
\newcommand{\ann}{{\rm ann}}
\makeatother

\begin{center}
\vglue .06in
{\Large \bf {Cosmological and astrophysical bounds on a\\ 
heavy sterile neutrino and the KARMEN anomaly}}
\bigskip
\\{\bf A.D. Dolgov
\footnote{Also: ITEP, Bol. Cheremushkinskaya 25, Moscow 113259, Russia.}
\footnote{e-mail: {\tt dolgov@fe.infn.it}},
S.H. Hansen\footnote{e-mail: {\tt sthansen@fe.infn.it}}
 \\[.05in]
{\it{INFN section of Ferrara\\
Via del Paradiso 12,
44100 Ferrara, Italy}
}}
\\{\bf G. Raffelt \footnote{e-mail: {\tt raffelt@mppmu.mpg.de}}} \\
{\it{Max-Planck-Institut f\"ur Physik (Werner-Heisenberg-Institut)\\
F\"ohringer Ring 6, 80805 M\"unchen, Germany
}}
\\{\bf D.V. Semikoz \footnote{e-mail: {\tt semikoz@ms2.inr.ac.ru}}} \\
{\it{Max-Planck-Institut f\"ur Physik (Werner-Heisenberg-Institut)\\
F\"ohringer Ring 6, 80805 M\"unchen, Germany\\
and\\
Institute of Nuclear Research of the Russian Academy of Sciences\\
60th October Anniversary Prospect 7a, Moscow 117312, Russia}}
\\[.40in]
\end{center}

\begin{abstract}

Constraints on the lifetime of the heavy sterile neutrino, that was 
proposed as a possible interpretation of the KARMEN anomaly, 
are derived from primordial
nucleosynthesis and SN~1987A. Together with the recent
experimental bounds on the $\nus$ lifetime, SN~1987A completely excludes
this interpretation. Nucleosynthesis arguments permit a narrow window
for the lifetime in the interval 0.1--0.2~sec. If $\nus$ possesses an 
anomalous interaction with nucleons, the SN bounds may not apply,
while the nucleosynthesis ones would remain valid.

\end{abstract}    


\newpage

\section{Introduction}

In 1995 the KARMEN collaboration discovered an anomaly in the time distribution
of the charged and neutral current events induced by neutrinos from $\pi^+$
and $\mu^+$ decays at rest~\cite{karmen95}. This anomaly may be explained by
the production of a new neutral particle in pion decay
\be
\pi^+ \rar \mu^+ + x^0~,
\label{piondecay}
\ee
with the mass 33.9 MeV, barely permitted by the phase space, so that this
particle moves with non-relativistic velocity. Its subsequent 
neutrino-producing 
decays could be the source of the delayed neutrinos observed in the
experiment. The anomaly was recently confirmed by the same 
group~\cite{karmen99} with better statistics and substantially reduced
cosmic-ray background. 

A search for $x^0$ particles produced in the rare pion 
decay (\ref{piondecay}) was performed in 1995 by an experiment at 
PSI~\cite{psi95}. It gave an upper limit for the branching
ratio of ${\rm BR}( \pi^+ \rar \mu^+ + x^0) < 2.6 \times 10^{-8}$ 
at 95\% CL.  

Several candidates for $x^0$ have been proposed in the literature.  In
ref.~\cite{barger95} the authors considered a sterile neutrino, $x^0
\equiv \nu_s$.  Their conclusion was that the sterile-neutrino
hypothesis is compatible with all laboratory constraints, but they
noted possible problems with astrophysics and cosmology.  Further
laboratory constraints on this model were investigated in
ref.~\cite{govaerts96}, concluding that mixings of $\nu_s$ with $\nue$
and $\num$ must be very small, while a mixing with $\nut$ was
permitted. In this case $\nu_s$ would predominantly decay through
neutral-current interactions into $\nut + \ell + \bar\ell$, where
$\ell$ is any light lepton, $\ell=\nue$, $\num$, $\nut$, or $e^-$.
The lifetime of $\nu_s$ with respect to this decay was estimated to be
in the range
\be
10^{-3}~{\rm sec}<\tau_{\nu_s} < 150~{\rm sec}~.
\label{taunus}
\ee
The lower bound on $\tau_{\nu_s}$, which comes from the experimental
bound on the $\nu_\tau$ mass (large mixing with $\nu_s$ makes 
it too heavy), was discussed in ref.~\cite{ahluwalia97}. 

In ref.~\cite{choudhury96} it was suggested that $x^0\equiv\tilde\chi$ 
could be the lightest 
supersymmetric particle, photino or zino, that decayed through the channel
\be
\tilde \chi \rar \gamma + \num~.
\label{chi2}
\ee
The new data~\cite{karmen99}, however, do not agree with the predictions of
the model so that recently a new version of supersymmetric model was 
considered~\cite{choudhury99}, according to which the light
neutralino decayed through a three body channel
\be
\tilde \chi \rar e^++e^-+\nu_{\mu,\tau}~.
\label{chi3}
\ee

In the paper~\cite{gninenko98} a new decay mode of muons was proposed as a 
source of the anomaly:
\be
\mu^+ \rar e^+ + S~,
\label{scalar}
\ee
where $S$ is a scalar boson with mass 103.9~MeV. A search for these decays
was reported 
in ref.~\cite{bilger99} where the upper limit 
${\rm BR}(\mu^+\rar e^+ + S) < 5.7\times 10^{-4}$ was obtained.
This limit, though it does not 
exclude the model, makes it more complicated. 

Recent data by the NOMAD collaboration~\cite{nomad} permit to 
strengthen the bound on the mixing of $\nu_s$ with $\nu_\tau$.
Their expected lower limit on $\tau_{\nu_s}$ is around
0.1~sec.

On the other hand, cosmology and astrophysics permit to obtain an upper
bound on $\tau_{\nu_s}$ that may be complementary to direct 
experiments. 
If this happens to be the case, then the explanation of the KARMEN 
anomaly by a 33.9 MeV sterile neutrino would be ruled out. Recently 
two papers have appeared~\cite{goldman99,kach99} where the observation of
SN~1987A were used to put bounds on the properties of the
proposed sterile neutrino or light neutralino. In the present paper we derive
constraints on the lifetime of a 33.9~MeV sterile neutrino from 
both more detailed consideration of SN~1987A and of primordial nucleosynthesis.
With the present-day accuracy of the data on the primordial light-element 
abundances, the bounds that are found from SN~1987A tend to be
stronger
than those found from primordial nucleosynthesis. However, the latter
remain interesting, first, because they may be competitive in the nearest 
future with an improved accuracy of the data on primordial nucleosynthesis,
and second, in a hypothetical 
case where $\nus$ would have an anomalously strong (stronger than the usual
weak) interaction with nucleons. In that case the
SN~1987A bounds would not apply, while the nucleosynthesis bounds would 
survive.

\section{Primordial Nucleosynthesis \label{ns}}

\subsection{General Features \label{gen}}  

A heavy unstable sterile neutrino would influence big-bang 
nucleosynthesis (BBN) through its contribution to the 
cosmological energy density by
speeding up the expansion and enlarging the frozen
neutron-to-proton ratio, $r_n=n/p$, and less directly, though stronger, 
through 
its decay products, $\nue,\, \num$, and $\nut$. The impact of $\num$ and
$\nut$ on BBN is rather straightforward: their energy density increases
with respect to the standard case and this also results in an increase
of $r_n$. This effect can be described by the increased number of
effective neutrino species $N_\nu$ during BBN (in the standard case
$N_\nu =3$). The increase of the energy density of $\nue$, due to decay of 
$\nus$ into $\nue$, has an opposite effect on $r_n$. Though a larger energy
density results in faster cooling, the increased number of $\nue$
would preserve thermal equilibrium between neutrons and protons for a longer 
time and correspondingly the frozen $n/p$-ratio would become smaller.
The second effect is stronger, so the net result is a smaller $n/p$-ratio. 
There is, however, another effect which is 
related to the distortion of the energy
spectrum of $\nue$ from the decays of $\nus$. If the spectrum is distorted
at the high-energy tail, as is the case, then proton formation in the
reaction $n+\nue \rar p + e^-$ would be less efficient than neutron
creation in the reaction $ \bar \nue + p \rar n + e^+$. We found that
this effect is quite significant. The decays of $\nus$ into the 
$e^+e^-$-channel will inject more energy into the electromagnetic part 
of the primeval plasma and this will diminish the relative contribution of
the energy density of light neutrinos and
diminish $r_n$.

We could use the technique and numerical code from our earlier 
papers where the effects of non-equilibrium massless neutrinos (in the
standard model)~\cite{dolgov97} 
and possibly massive~\cite{dolgov98} and/or unstable~\cite{dolgov99} 
tau-neutrinos were precisely calculated.
However, since we do not need the accuracy of a fraction of per cent achieved
in these papers, we will use a simpler and considerably less time consuming
approximate approach that we will now discuss.

\subsection{The model}
\label{model}

If the KARMEN anomaly is explained by 
 a heavy sterile neutrino with $m_{\nu_s}=33.9$~MeV then, as
was mentioned in the Introduction, it may only mix with $\nut$,
\begin{eqnarray}
\nu_\tau &=& \nu_1 \, \cos\Theta  + \nu_2 \,\sin\Theta ~,
\nonumber \\
\nu_s &=& -\nu_1 \, \sin\Theta  + \nu_2 \, \cos\Theta ~,
\label{vac_oscil}
\end{eqnarray}
where $\Theta$ is the vacuum mixing angle and
$\nu_1$ and $\nu_2$ are the mass eigenstates; the mass difference is
positive: $\delta m^2=m_2^2-m_1^2\approx (33.9~{\rm MeV})^2 > 0$.
Through neutral-current interactions $\nus$ could decay into $\nut$
and a pair of other light leptons. 
The corresponding processes and their matrix elements are presented in 
table 1. The lifetime of $\nus$ is given by the expression
\be
\tau_{\nus} \equiv \Gamma_{\nus}^{-1} =
\left[ {(1+\tilde g_L^2 +g_R^2) G_F^2 m_{\nus}^5 (\sin^2 2\Theta)/4 \over
192 \pi^3 }\right]^{-1} 
\approx {5.7\times 10^{-4}\,{\rm sec}\over(\sin^2 2\Theta)/4}~,
\label{gammas}
\ee
where
\be
\tilde{g}_L = -1/2 + \sin^2 \theta_W\,\, {\rm and}\,\, g_R = \sin^2 \theta_W~.
\label{glgr}
\ee

According to the combined experimental data, the lifetime lies in the
range
\be
0.1~{\rm sec} < \tau_{\nus}< 150~{\rm sec}
\label{exp_bounds}
\ee
and correspondingly
\be
3.9\times 10^{-3} < \sin 2\Theta <0.15~.
\label{thetalim}
\ee
Even with a very small mixing, $\nus$ could be abundantly produced in the 
early universe when the temperature was higher than $\mnus$.
Their production rate can be estimated as~\cite{barbieri90,enqvist92}
\begin{equation}
\Gamma_{\nus} = \frac{1}{2} \sin^2 2 \Theta_\M \Gamma_{W}, 
\label{Gamma}
\end{equation}
where $\Gamma_W= 2.5\, G_F^2 T^5$ is the averaged weak interaction rate
and $\Theta_\M$ is the mixing angle in the medium. 
According to the calculations of
refs.~\cite{barbieri90,notzold88}
\be
\sin 2 \Theta_\M \approx {(\sin 2\Theta)/2 \over
1+ 0.76 \times 10^{-19}\, T^6 (\delta m^2)^{-1}}
\approx  {(\sin 2\Theta)/2 \over
1+ 6.6\times 10^{-23}\, T^6 }~,
\label{thetam}
\ee
where $T$ and $\delta m^2$ are taken in MeV. One sees that matter effects 
are not important for $T < 5$~GeV. 

Comparing the production rate (\ref{Gamma}) with the Hubble expansion rate
\begin{equation}
H = \sqrt{\frac{8 \pi^3}{90} g_*(T)}\, \frac{T^2}{M_{\rm Pl}},
\label{hubble}
\end{equation}
we find that sterile neutrinos could be abundantly produced in the early
universe if, roughly speaking, $(\sin 2\Theta)^2 (T/3~{\rm MeV})^3 >1$.   
Even for $\Theta$ at the lower limit given by the relation (\ref{thetalim}),
the equilibrium condition is fulfilled for $T\approx 120$~MeV. For this small
value of $\Theta$ the original equilibrium number density of $\nus$ would be
diluted at smaller $T$ by annihilation of pions and muons. However, as we see
in what follows, nucleosynthesis strongly disfavors large values of $\taus$.
Correspondingly, for large mixing angles $\nus$ remains in equilibrium 
for much smaller $T$, and this dilution is not essential.

The evolution of $\nus$ at lower temperatures,
$T<m_{\nus}$, is considered below. We calculate the number density of 
the heavy $\nus$ assuming that initially they were in 
equilibrium  and in the process of freeze-out they interacted with
the thermal equilibrium bath of light particles. 
With the evident correction for the
$\nus$ decay, these calculations are very similar to the usual freeze-out
calculation of massive species. There is one important difference
however. Normally massive particles disappear in the process of mutual 
annihilation, so that the rate of freezing is proportional to the number
density of the particle in question, $\dot n_m /n_m \sim \sigma_{\ann}n_m$,
which becomes exponentially small when $T<m$. Sterile neutrinos may also 
disappear in collisions with massless leptons
\be
\nus + \ell_1 \rar \ell_2 + \ell_3~,
\label{nusl}
\ee
so their
extinction through this process would be more efficient than by mutual
annihilation, $\nus +\bar\nus \rar {\rm all}$. The decoupling temperature of the 
process (\ref{nusl}) can be approximately found from the decoupling of the 
usual massless neutrinos at $T_{\nut}\sim 2$~MeV by rescaling it by the 
mixing parameter, $(\sin 2 \Theta)^{2/3}$. To obtain a better estimate
we solve the corresponding kinetic equation for $\nus$ in sec.~\ref{numnus}.
This permits us 
to determine the distribution function, $f_{\nus} (x,y)$, where
\be
x =  m_0 a(t)\quad {\rm and}\quad y = p\, a(t)~,
\label{xy}
\ee
are convenient dimensionless variables with 
$a(t)$ the cosmic scale factor
and $m_0$ the normalization mass that we choose as $m_0 = 1$~MeV.
In what follows we will often skip $m_0$, keeping in mind that the
relevant quantities are measured in MeV.
In terms of these variables, the kinetic equations have the form
\be
(\partial_t - Hp\partial_p ) f = Hx\partial_x f = I_{\coll}~,
\label{kineqxy}
\ee
where
\be
I_{\coll} &=& {1\over 2E} \int \prod_i \left( {d^3 p_i \over 2E_i (2\pi)^3}
\right)
\prod_f \left( {d^3 p_i \over 2E_f (2\pi)^3} \right) \nonumber \\
&&{}\times (2\pi)^4 \delta^{(4)} (\sum_i p_i -
\sum_f p_f) \left| A_{if} \right|^2 F(f_i,f_f)~,
\label{icoll}
\ee
is the collision integral and 
\be
F(f_i,f_f) = - \prod_i f_i \prod_f (1-f_f) +\prod_f f_f \prod_i (1-f_i)
\label{foff}
\ee
with sub-$i$ and sub-$f$ meaning initial and final particles.

After the distribution function of $\nus$ is found, 
the next step is to find the distribution functions of the light neutrinos and
in particular 
their energy densities. The distributions of electrons and positrons are
of course assumed to be very close to equilibrium because of their very 
fast thermalization due to the interaction with the photon bath. However, the
evolution of the photon temperature, due to decay and annihilation of the
massive $\nus$, becomes different from the standard one, $T_{\gamma} \sim 1/x$,
by an extra factor $(1+\Delta) >1$:
\be
T_{\gamma} = [1+ \Delta (x)] /x~.
\label{tgamma}
\ee
At sufficiently high temperatures, $T>T_W \sim 2$~MeV,
light neutrinos and electrons/positrons were in strong contact, so that
the neutrino distributions were also very close to the equilibrium ones. If
$\nus$ disappeared sufficiently early, while 
thermal equilibrium between $e^{\pm}$ and neutrinos remained, then 
$\nus$ would not have any observable effect on primordial abundances, 
because only the contribution of neutrino energy density relative to 
the energy density of $e^{\pm}$ and $\gamma$ is
essential for nucleosynthesis. Hence a very short-lived $\nus$ has a negligible
impact on primordial abundances, while with an increasing lifetime the
effect becomes stronger. Indeed at $T<T_W$ the exchange of energy between 
neutrinos and electrons becomes very weak and the energy injected into the 
neutrino component is not immediately redistributed between all the particles.
The branching ratio of the decay of $\nus$ into $e^+e^-$ is approximately
1/9, so that the neutrino component is heated much more than the 
electromagnetic one. As we mentioned above, this leads to a faster cooling
and to a larger $n/p$-ratio. 

So, for the numerical calculations  we adopt the following procedure. 
We assume that at sufficiently high
temperature, e.g.~$T_i=5$ MeV, or equivalently 
at the initial value of the scale factor
$x_{i} = 1/T_i = 0.2$, there is a complete thermal equilibrium between 
active neutrinos and electrons. Starting from that moment we calculate the 
corrections to the active neutrino distribution functions
\be
\delta f_{\nu} = f_\nu - f^{\eq}_\nu~,
\label{deltaf}
\ee
where the equilibrium distribution function is assumed to have the standard
Fermi-Dirac form with a temperature that drops  as $1/x$,
\be
f^{\eq}_\nu = (e^y +1)^{-1}~.
\label{feq}
\ee
The evolution of the photon temperature, i.e.~$\Delta (x)$, is determined 
from the energy balance equation
\be
d\rho / dx = -3 (\rho + p)/x~,
\label{drhodx}
\ee
where $\rho$ and $p$ are respectively the total energy and pressure densities
in the cosmological plasma. 

It is worth noting that we normalized the scale factor in such a way that
at the initial moment $x_i T_i =1$. If we change the initial moment 
for calculating $\Delta$, it would 
result in a different definition  of $x$, but this is unobservable. 
We have checked that for $x_i = 0.1$--0.3 the results weakly depend upon the
value of $x_i$, and that the corrected neutrino distribution 
$f_\nu = f_\nu^{\eq} +\delta f$ quite accurately maintain
the equilibrium shape with the same temperature as 
electrons, $T_{\nu}=T_\gamma$ at these early times.

We made several assumptions that permitted to simplify calculations
very much: $\nus$ was assumed non-relativistic, the equilibrium
distribution functions ``inside'' the collision integral were taken in
the Boltzmann approximation, while ``outside'' they were taken in
Fermi-Dirac form, and the kinetic equations for $\delta f_\nu$ were
taken in a simplified form, so that some fraction of neutrino energy
was lost (see sec.~\ref{lightnu}). All these assumptions lead to a
weaker impact of $\nus$ on nucleosynthesis, so the real bound should
be somewhat stronger than what is presented below.
 
One more comment is in order. We do not take into account oscillations
between $\nus$ and $\nut$ for $T<m_{\nus}$.  This is perfectly
justified because in the interesting range of neutrino
energies the oscillation frequency is so high and the velocities of
$\nus$ and $\nut$ are so much different that the coherence is quickly
lost and they can be considered as independent particles.  Medium
effects are also not important for the considered positive mass
difference $\delta m^2 \approx 10^3~{\rm MeV}^2$.

\subsection{Evolution of Heavy Neutrinos \label{numnus}}

The evolution of the occupation number of $\nus$ is determined by its decays
and inverse decays, listed in table~1, and by the
reactions~(\ref{nusl}) with all possible sets
of light leptons permitted by quantum numbers,  presented 
in table~2. 
Taking both contributions
into the collision integral and assuming that the massless species are in
thermal equilibrium with temperature $T$ and that both helicity 
states of $\nus$ are equally populated, we obtain
\be
\partial_x f_{\nus} (x,y) &=& {1.48\, x \over \tau_{\nus}({\rm sec})} 
\left( {10.75 \over g_* (T) }\right)^{1/2} 
{f_{\nus}^{\eq} - f_{\nus} \over (Tx)^2} \nonumber \\
&& \left[ {m_{\nus}\over E_{\nus}} 
+ {3\times 2^7 T^3 \over \mnus^3 } \left( \frac{3 
\zeta(3)}{4}
+ \frac{7 \pi^4}{144} \left( {E_{\nus} T\over \mnus^2} + 
{p_{\nus}^2 T \over 3 
E_{\nus}\mnus^2} \right) \right) \right]~,
\label{dxfs}
\ee 
where $E_{\nus}=\sqrt{\mnus^2 +(y/x)^2}$ and $p_{\nus}= y/x$ are the
energy and momentum of $\nus$ respectively and $g_* (T)$ is the
effective number of massless species in the plasma determined as the
ratio of the total energy density to the equilibrium energy density of
one bosonic species with temperature $T$, $g_* = \rho_{\tot} /(\pi^2
T^4 /30)$.  The coefficient $1.48$ comes from expressing the lifetime
in sec according to the relation $1/{\rm MeV} = 0.658\times
10^{-21}$~sec and from the value of the Hubble parameter
$H=\sqrt{8\pi\rho_{\tot}/3M_{\rm Pl}^2}$.  The first term in square
brackets, $m_{\nus}/E_{\nus}$, comes from summing the squared matrix
elements in table~1 (decay) while the rest is obtained by using that
the sum of the squared matrix elements from table~2 (collisions) can
be written as
\be
|M|^2 \sim \left( 1+\tilde g_L^2 + g_R^2 \right) \, 
\left[ (p_1\cdot p_2)(p_3\cdot p_4) + 
2 (p_1\cdot p_4)(p_3\cdot p_2) \right] ~.
\label{reactions2}
\ee

The assumption of Boltzmann statistics for massless leptons considerably 
simplifies the calculations, because the integral over their 
momenta can be taken explicitly. This approximation results in a larger
value of the collision integral, i.e.~in faster decay and reaction rates
and correspondingly to a smaller abundance of $\nus$. Thus the restriction
on $\taus$ obtained in this approximation is weaker than the real one.

At high temperatures, $T> 5$~MeV, (\ref{dxfs}) was integrated with the
simplifying assumption $Tx =1$. This also leads to a weaker bound on
$\taus$. Assuming entropy conservation one can check that $Tx$ may
change by a factor 1.1 due to $\nus$ decays and annihilation. In fact
the effect is stronger because the number density of $\nus$ is much
larger than the equilibrium one. However, since we started at
$T=5$~MeV $<m_{\nus}$, when the energy density of $\nu_{\nus}$ was
already somewhat suppressed, the variation of $Tx$ from that moment
would be weaker.

Naively one would expect a
variation of $f_{\nus}$ similar to the variation of $Tx$. However, it is
argued below that the effect may be much stronger.
Indeed, since $\nus$ disappears in the collisions with massless particles,
its number density is much more sensitive to the variation of the 
coefficient in front of the collision integral. In the case of the
standard freezing in two-body annihilation the frozen number 
density, $n_f$, is known 
to be inversely proportional to the annihilation cross-section.
This is not true for the case of reactions (\ref{nusl}). One can solve the 
kinetic equation explicitly and check that the result is
exponentially sensitive to the coefficient in front of 
$(f_{\nus}^{\eq} - f_{\nus})$. Therefore
we took the variation of $Tx$ into account, starting from $x_{i}=0.2$, 
according to the energy balance law (\ref{drhodx}),
see also (\ref{ddeltadx}). 
A simpler and more common method based on
entropy conservation is not accurate 
enough because the sterile neutrinos strongly deviate from equilibrium in 
the essential range of temperatures and thus entropy is not conserved.

In the non-relativistic limit we may integrate both sides of
(\ref{dxfs}) over $d^3y /(2\pi)^3$ to obtain the number density of
$\nus$ in the comoving volume
\be
\bar n_{\nus} = x^3 n_{\nus}=\bar n_{\nus}^{\eq} - \int_0^x dx_1 
{d\bar n_{\nus}^{\eq}\over dx_1} 
\exp\left[\int_{x_1}^x dx_2 B( x_2) \right] ~,
\label{fnussol}
\ee  
where
\be
B(x) =  {1.48 x \over (Tx)^2 (\tau_{\nus}/{\rm sec})} 
\left( {10.75 \over g_* (T) }\right)^{1/2} 
\left[ 1 + {3\times 2^7 (Tx)^3 \over x^3 \mnus^3 } \left( \frac{3 
\zeta(3)}{4}
+ { \frac{7 \pi^4}{144} (Tx)\over x \mnus } \right) \right]~.
\label{bx}
\ee
Usually $Tx$ is assumed to be constant, quite 
often normalized as $Tx =1$. A small variation of this quantity usually
is not very important for the value of $\bar n$. In our case, however,
estimating the integral by the saddle point method reveals that
the dependence on $Tx$ appears in the exponent and the effect may be
significant even for a small variation of $Tx$. The dependence on
$Tx$ for the scattering term results in an increase of $\bar n_{\nus}$,
while that for the decay term results in a decrease of $\bar n_{\nus}$.

\begin{figure}[bt]
\psfig{file=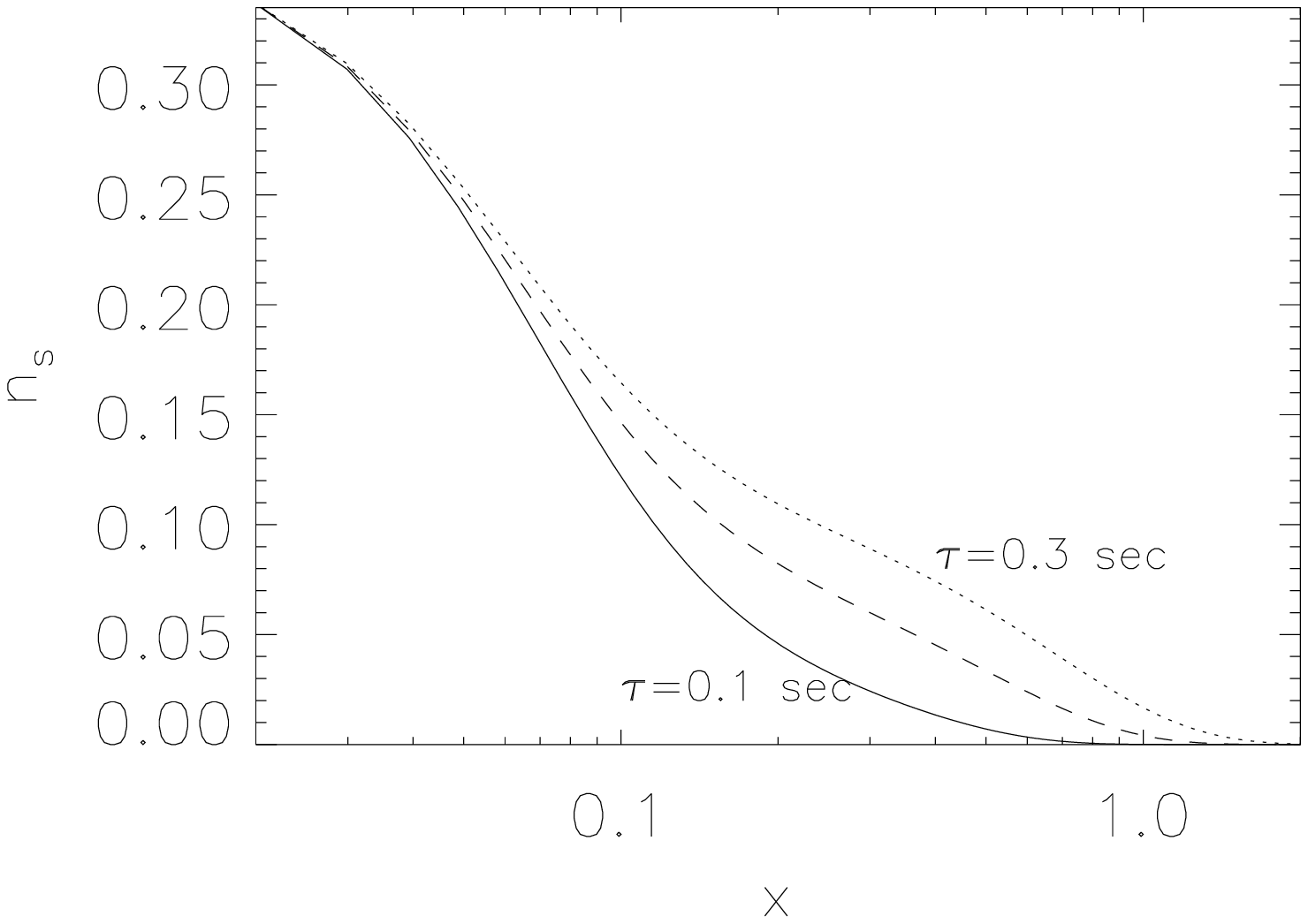,width=4.5in,height=3.2in}
\caption{The normalized number density, $x^3\, n_{\nus}$, as a function of 
$x$ for the lifetimes 0.1~sec (solid), 0.2~sec (dashed) and 0.3~sec (dotted).}
\label{fig:ns_of_x}
\end{figure}

\begin{figure}[hbt]
\psfig{file=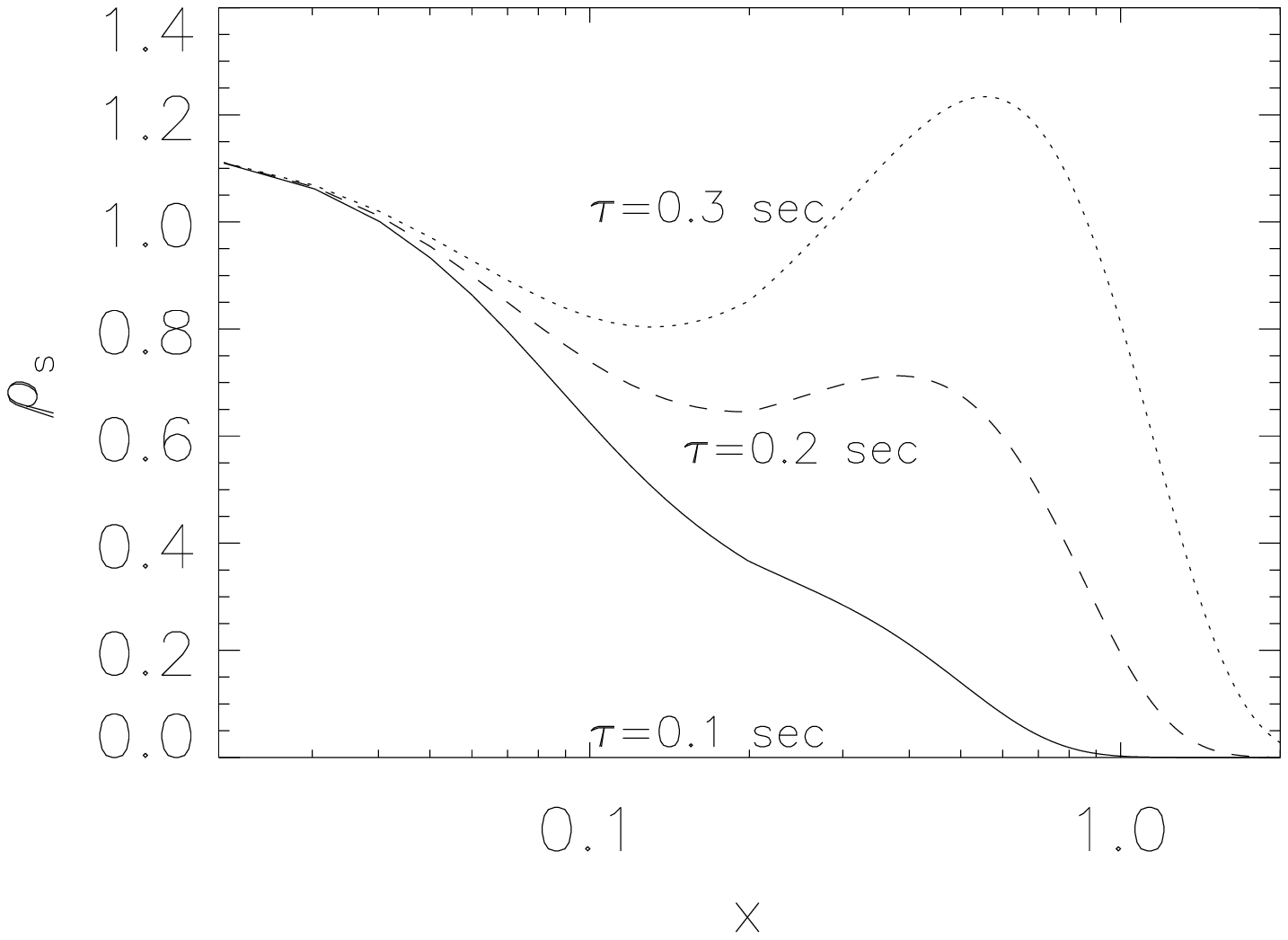,width=4.5in,height=3.2in}
\caption{The normalized energy density, $x^4\, \rho_{\nus}$, 
as a function of $x$ for the
lifetimes 0.1~sec (solid), 0.2~sec (dashed) and 0.3~sec (dotted).}
\label{fig:rhos_of_x}
\end{figure}

\begin{figure}[hbt]
\psfig{file=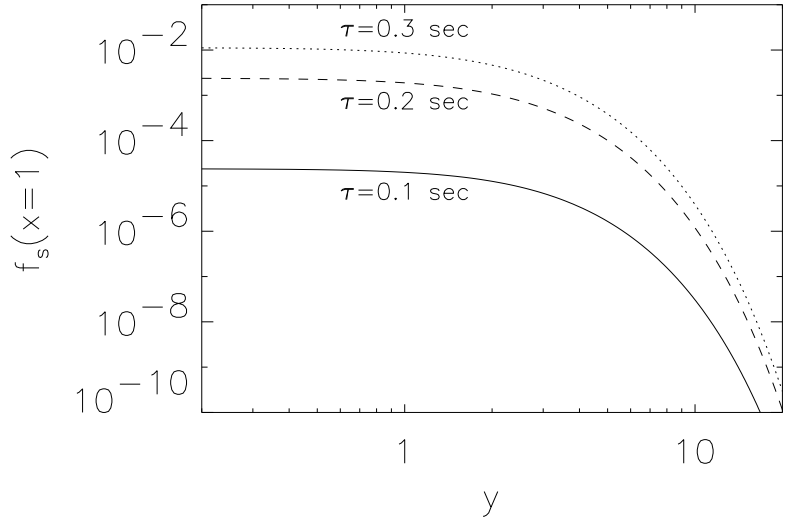,width=4.5in,height=3.2in}
\caption{A snap-shot of the distribution function, $f_{\nus}(y)$, 
at the time $x=1$ for the
lifetimes 0.1~sec (solid), 0.2~sec (dashed) and 0.3~sec (dotted).}
\label{fig:fs_of_y}
\end{figure}

The results of the numerical solution of (\ref{dxfs}) for different
values of lifetimes are presented in figs.~\ref{fig:ns_of_x},
\ref{fig:rhos_of_x}, and \ref{fig:fs_of_y}.  The first one shows the
evolution of $x^3 n_{\nus}$ for $\taus = 0.1,\, 0.2,\,\, {\rm
and}\,\,0.3$, the second is $x^4 \rho_{\nus}$, while the third figure
is a snap-shot of the distribution functions, $f_{\nus}$, at $x=1$.
All the distributions are significantly higher than the equilibrium
ones. For example, the equilibrium distribution is about 10 orders of
magnitude below the calculated curve for $\tau_{\nus}=0.1$.

\subsection{ Light Neutrinos \label{lightnu}}

At the onset of nucleosynthesis, $\nus$ practically disappeared
from the primeval plasma, at least for sufficiently small lifetimes, which
are near the bound obtained below. However, the products of their decays
and annihilation distorted the standard nucleosynthesis conditions:
the energy density as well as the spectrum of neutrinos were 
different from the standard ones, and this changed the
light element abundances. 

We will look for the solution of the 
kinetic equations governing the evolution of the distribution
functions of light neutrinos in the form
\be
f_{\nu_a}(x,y) = \left( 1+ e^{y} \right) ^{-1} + \delta f_{\nu_a}(x,y)~,
\label{fnua}
\ee
where $a = e,\,\,\mu,\,\, {\rm or}\,\, \tau$.
The first term is equal to the equilibrium distribution function for the case
when the 
temperature drops as the 
inverse scale factor, i.e.~$Tx =1$. In reality 
this is not the case and the function $\delta f$ takes into account both the
variation of $Tx$ and the spectrum modification of the active neutrinos. 

The distribution of $e^{\pm}$ always has the equilibrium form, 
$f_e = [1+\exp (E/T)]^{-1}= [1+\exp (y/Tx)]^{-1} $ 
and the product $Tx$ is taken in the
form
\be 
Tx = 1 + \Delta (x)~,
\label{deltax}
\ee
where $\Delta$ is assumed to be small, such that a perturbative expansion in 
$\Delta$ can be made. A similar 
first-order perturbative expansion is made with 
respect to $\delta f$, so that the collision integral becomes linear in terms
of $\Delta$ and $\delta f$. 

It is assumed that initially both $\delta f$ and $\Delta$ were
zero. To this end one should find an appropriate value of the initial
``time'', $x_i$, or temperature, $T_i$. The temperature should be
sufficiently small, such that the number density of $\nus$ is already
low and hence the function $\Delta$ would not rise too much. On the
other hand, the temperature should not be too low, otherwise
equilibrium between neutrinos and electrons would not be established
and the assumption of $\delta f=0$ could be grossly wrong. As we see
in what follows, convenient initial values that satisfy both
conditions are $x_i \sim 0.2$ and $T_i\sim 5$~MeV for $\taus \sim 0.2$
sec. Of course, the later rise of temperature by the well known factor
1.4 because of $e^+e^-$-annihilation is taken into account explicitly.

\subsection{Collision Integral and Source Term}
\label{collsource}

There are two kinds of terms in the collision integral, ``easy'' ones where
the unknown function $\delta f$ depends upon the external variable $y$,
which is not integrated upon, and ``difficult'' ones when $\delta f$ is 
under  the integral. The terms of the first kind come, with
negative sign, from external particles in the initial state. They give
\be
\partial_x \delta f_{\nue} (x,y) = - 0.26 \left({10.75 \over g_*}
\right)^{1/2} \delta f_{\nue}(1+{g}_L^2+g_R^2) 
(y/ x^4) ~,
\label{dxdeltaf}
\ee
where $g_L = \sin^2 \theta_W +0.5$, while $g_R =\sin^2 \theta_W$.
For $\nu_\mu$ and $\nu_\tau$ the  functions $\tilde g_L$ and $g_R$  are given 
by expressions (\ref{glgr}).
Using the simple expression (\ref{dxdeltaf}) one finds that the 
rate of approach of $\nue$
to equilibrium is given by
\be
\delta f_{\nue} \sim \exp (-0.13\, y/x^3)~.
\label{deltaf1}
\ee
One sees that equilibrium would be efficiently restored for $\nue$ if
$x<0.5\,y^{1/3}$ or $T>\,2 y^{-1/3}$~MeV, while for $\nu_{\mu,\tau}$
equilibrium would be restored if $T> 2.25\, y^{-1/3}$~MeV. These
results are close to the standard estimates known in the literature
(for the most recent ones see e.g.~\cite{kimmo}). However, there are
the following drawbacks in this derivation. First, it was done under
the assumption that the source of distortion acted for a finite time
and thus (\ref{deltaf1}) could be valid only after the source has been
switched off.  If there is a constantly working source, $S(x,y)$,
equilibrium is always distorted roughly by the factor $S/\Gamma$,
where $\Gamma$ is the effective reaction rate. Another and more
serious argument against the validity of (\ref{deltaf1}) is that the
``difficult'' part of the collision integral was neglected. One can
see from the general expressions~(\ref{icoll},\ref{foff}) that $\delta
f$ also appears under the integral over momenta, and it can be seen
that it comes mostly with a positive sign. These terms counteract the
smoothing action of expression~(\ref{dxdeltaf}) and shift equilibrium
restoration to considerably higher temperatures. One can take them
into account exactly, numerically solving the kinetic equations with
the exact collision integral that can be reduced down to a
2-dimensional integration over particle energies (see
e.g.~\cite{dolgov97}), but this is a very time consuming
procedure. Instead we will here use a much simpler approach.  We will
approximately represent such ``difficult'' terms by the integrals over
energy chosen in such a way that the kinetic equations satisfy the law
of particle conservation in the comoving volume if only elastic
scattering is taken into account, i.e.~the kinetic equation should
automatically give $\partial_x (x^3 n) = 0$.  The exact equations
should simultaneously satisfy energy conservation law, but working
with the exact equations is much more complicated.  We checked the
validity of our approximate procedure for a case of a more simple
reaction amplitude, where we compared the approximate and exact
results and found very good agreement. The approximation that we use
breaks the energy conservation law in the kinetic equation, so that
some small fraction of energy going into light particles from $\nus$
decays and/or annihilation is lost. This diminishes the effects that
we are discussing so that the real constraints on $\taus$ should be
stronger.

The kinetic equations for light neutrinos in this approximation can be
written as
\be
\partial_x \delta f_{\nue} (x,y) &=& S_{\nue}(x,y) +
0.26 \left({10.75 \over g_*}
\right)^{1/2} (1+g_L^2+g_R^2) (y/ x^4)\\
&&{}\times\biggl\{- \delta f_{\nue} 
+{2\over 15}\,{e^{-y}
\over 1+g_L^2+g_R^2} \, [1 +0.75(g_L^2+g_R^2)]\nonumber\\
&&\hskip2em{}\times
\left[ \int dy_2 y^3_2 \delta f_{\nue} (x,y_2)
+{1 \over 8}\, 
\int dy_2 y^3_2 \left( \delta f_{\num} (x,y_2)+\delta f_{\nut}(x,y_2)
\right) \right] \nonumber \\
&&\hskip2em{}+ {3\over 5}\Delta (x){g_L^2+g_R^2 \over 1+g_L^2+g_R^2} e^{-y}
\left( 11y/12 -1\right) \biggr\}~.\nonumber
\label{dxfnuall}
\ee
The coefficients $g_{L,R}$ for $\nu_{\mu,\tau}$ are given by 
(\ref{glgr}) and
for $\nue$ are presented after (\ref{dxdeltaf}). 
The source term $S$ describes injection of non-equilibrium
neutrinos by $\nus$ decays or reactions (\ref{nusl}) with light leptons. 
In what follows we include only decays. We have estimated the contributions
of the reactions~(\ref{nusl}) and 
found that they slightly improve the restrictions that
we have obtained.
For $\nue$, $\num$, and $\nut$ the contributions of the decay term
are respectively
\be
S_{\nue,\num}(x,y) &=& {0.012 \over \taus x^2 }\left({10.75 \over g_*}
\right)^{1/2}
\left( 1-{16y\over 9 m_{\nus}x}\right) \left( n_{\nus} - n_{\nus}^{\eq} \right)
\theta \left(m_{\nus}x/2-y\right) ~,
\label{snuem} \\
S_{\nut}(x,y)&=&{0.024 \over \taus x^2 }\left({10.75 \over g_*} \right)^{1/2}
\left[ 1-{16y\over 9 m_{\nus}x} +{2\over 3}\left(1+{\tilde g}_L^2 +g_R^2\right)
\left(1 - {4y \over 3 m_{\nus}x} \right)
\right] \nonumber \\
&&{}\times 
\left( n_{\nus} -n_{\nus}^{\eq}\right) \theta \left(m_{\nus}x/2-y\right)~,
\label{snut}
\ee
where $n_{\nus}(x)$ is the number density of $\nus$ and $\theta (y)$
is the step function which ensures energy conservation in the decay.
The factor $0.012$ comes from the product of the branching ratio
BR${}=96/4/(1+\tilde g_L^2+g_R^2) = 21.3$ with the factors $
3/m_{\nus}^3=7.7\times10^{-5}$ ($m_{\nus}$ in MeV) from the
normalization, we divide by $2/\pi^2=0.203$ 
from the number density, and 6.582 from
the relation between MeV and sec. Dividing by the Hubble parameter gives
a factor,
$1.221 /(1.88 \sqrt{8 \pi/3} ) = 0.2244$, and we find $21.3 \times
(7.7\times10^{-5})\times 0.2244 \times 6.582/0.203 =0.012$.

The coefficient in front of the collision integral is in reality momentum
dependent, and hence is slightly different from 0.4 for $\nu_e$ (and 0.29 for
$\nu_{\mu, \tau}$), which are often used in the literature. We extracted the
correct momentum dependent coefficients from our Standard Model 
code~\cite{dolgov97}, and 
used this in the calculations. This slightly weakens the bound on 
the lifetime.

The function $\Delta (x)$ is determined from the energy balance 
condition~(\ref{drhodx}) which in the present case reads
\be
{d\Delta \over dx} = -{1 \over 4 x^4 \rho_{EM} } \left[ { x d(x^3 \rho_{\nus})
\over dx} + {d (x^4 \delta \rho_{\nu}) \over dx} \right]~,
\label{ddeltadx}
\ee
where $\rho_{EM}$ is the energy density in the electromagnetic sector, and
$\rho_{\nus}$ and $\rho_{\nu}$ are the energy density of $\nus$ and
the total energy density of all light neutrinos respectively.
We also used another method for calculating $Tx$ starting
from earlier times and obtained stronger results, so we believe that 
the limits that we obtain here are quite safe.

\subsection{A Few Numerical Technicalities}

Let us summarize a few technicalities related to the numerical approach.  
We divide the time into 3 regions. First we integrate only 
(\ref{dxfs}) from very high temperatures, $T=50$ MeV ($x=0.02$)
and until $T=5$ MeV ($x=0.2$). We assume that initially the sterile particles
are in equilibrium (see sec.~\ref{model}). 
In this way we can follow the freeze out
and initial decay of the sterile particles, without worrying about the
equilibrium active neutrinos and the electromagnetic plasma. In the next
region, $0.2 < x < 50$, we solve (\ref{dxfs}, \ref{dxfnuall},
\ref{ddeltadx}) with the
initial values $\Delta=0$ and $\delta f_{\nu_{\rm active}}=0$. 
It is worth mentioning
that from the very beginning we separately calculate and include the
annihilation of the electrons, which increase the photon temperature with
a standard factor $1.4$. Finally, for very high $x>50$ ($T<0.03$ MeV) when 
all sterile neutrinos have disappeared and the active neutrinos have long 
decoupled, we solved only the kinetic equations governing the 
$n$-$p$-reactions 
needed for the nucleosynthesis code.
We use an 800 point grid in momentum in the region $0<y<80$, and checked
that the results are insensitive to the doubling of the grid.
For the BBN calculations we use $\eta_{10}=5$.

\subsection{Results}
\label{results}

We have solved  (\ref{dxfs}, \ref{dxfnuall}, \ref{ddeltadx}) numerically
for different lifetimes. In fig.~\ref{fig:ns_of_x} we plot
the evolution of the normalized number density, $x^3\, n_{\nus}$, as a function
of $x$. One sees for $\taus=0.3$ how $\nus$ first freeze out, followed by the 
subsequent decay. In fig.~\ref{fig:rhos_of_x} a similar plot of the
normalized energy density, $x^4 \rho_{\nus}$, is presented as a 
function of $x$,
and the effect of the sterile particle being massive is evident.
In fig.~\ref{fig:fs_of_y} we present a snap-shot of the distribution function 
at the time $x=1$. The equilibrium distribution function at this time is
about 10 orders of magnitude smaller than the curve for $\tau=0.1$~sec.

\begin{figure}[hbt]
\psfig{file=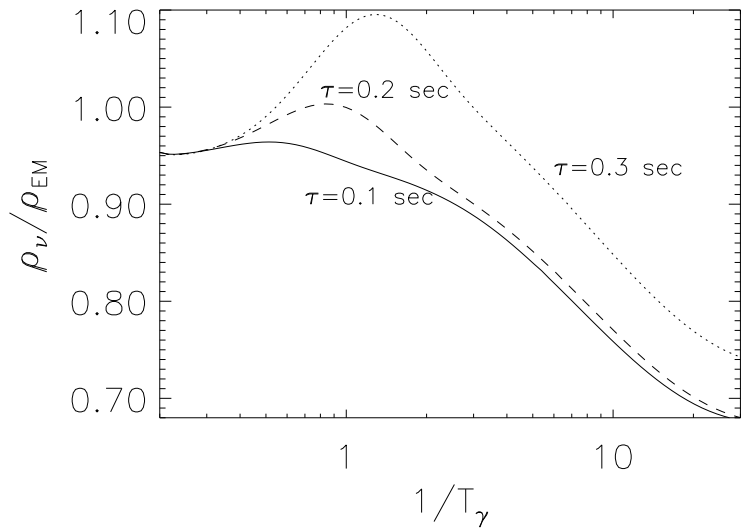,width=4.5in,height=3.2in}
\caption{The energy density of all the active neutrinos divided by the 
energy density in the electromagnetic plasma, $\sum \rho_{\nu}/\rho_{EM}$, 
as a  function of $x$
for lifetimes 0.1~sec (solid), 0.2~sec (dashed) and 0.3~sec (dotted).}
\label{fig:frac_of_x}
\end{figure}

\begin{figure}[hbt]
\psfig{file=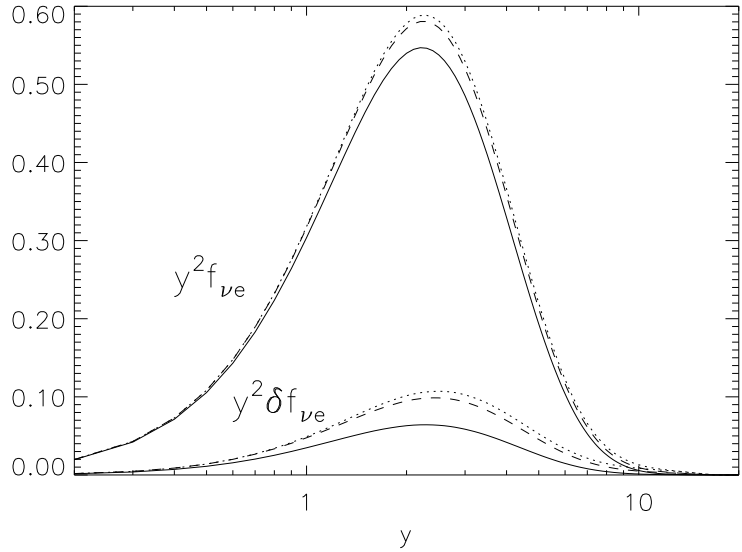,width=4.5in,height=3.2in}
\caption{A snap-shot of the spectrum of $\nue$, namely $y^2 f_{\nu_e}$ and 
the distortion $y^2 \delta f_{\nu_e}$ at $x=1$. The lifetimes
are 0.1~sec (solid), 0.2~sec (dashed) and 0.3~sec (dotted).}
\label{fig:fnue_of_y}
\end{figure}

The calculated energy densities
of all light neutrino species relative to the electromagnetic energy density,
$\rho_e + \rho_\gamma$, are presented in fig.~\ref{fig:frac_of_x} 
as functions of $x$. One sees that this fraction is higher for longer 
lifetimes, especially around the time $x=1$, when the $n/p$-ratio freezes
out, leading to an expected increase in the final helium abundance.
In fig.~\ref{fig:fnue_of_y} a snap-shot of
the spectrum of $\nue$, namely $y^2 f_{\nu_e}$ and  the distortion $y^2 \delta 
f_{\nu_e}$ are presented for $x=1$ for the lifetimes $0.1, 0.2, 0.3$. A 
distortion of the electronic neutrino spectrum has a strong impact on 
nucleosynthesis, while $\num$ and $\nut$ act only by their total energy
density. It is noteworthy that the increase in $\rho_{\nue}$ acts in the
opposite direction to an increase in $\rho_{\num,\nut}$, since it reduces
the effective number of light neutrinos, or in other words, it gives rise 
to a smaller mass fraction of primordial $^4$He, while an increase in the 
high energy part of $\nue$ spectrum results in a larger mass  fraction
of $^4$He.  

\begin{figure}[hbt]
\psfig{file=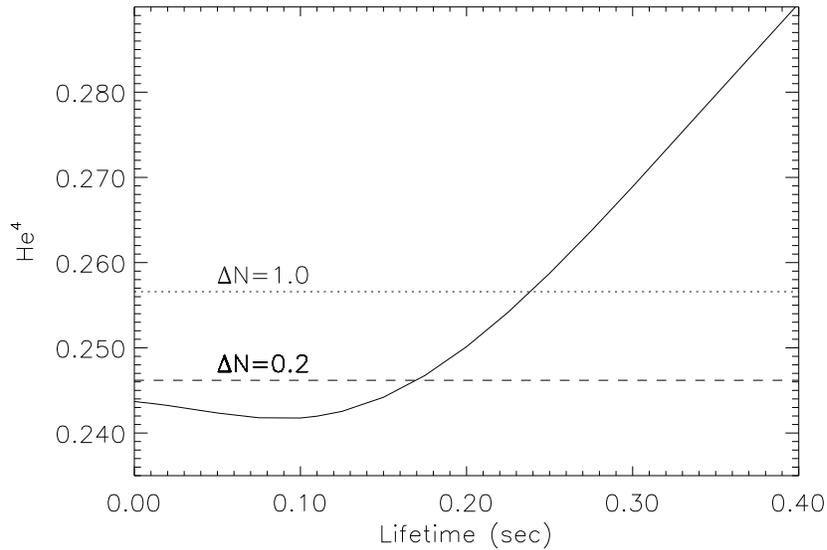,width=4.5in,height=3.2in}
\caption{The final helium abundance as a function of lifetime. The horizontal 
lines correspond to $\Delta N=0.2$ and $1.0$ extra effective neutrino species.}
\label{fig:he_of_tau}
\end{figure}

The results of the calculations have been imported into the modified
Kawano~\cite{kawano} nucleosynthesis code, and the abundances of all
light elements have been calculated.  At each time step $x$, we find
the corresponding photon temperature and total energy
density. Furthermore we integrate the kinetic equation governing the
$n/p$ evolution taking into account the distorted spectrum of $\nue$.
The final helium abundance is presented as a function of the $\nus$
lifetime in fig.~\ref{fig:he_of_tau}.  By translating these results
into effective number of neutrinos one sees that if we allow for
$\Delta N = 0.2$ (as suggested by~\cite{tytler0001318}), then only
lifetimes lower than $\tau_{\nus} = 0.17$ sec are permitted. If one is
more conservative and allows for one extra neutrino species, $\Delta N
= 1.0$, then lifetimes longer than $\tau_{\nus} = 0.24$ sec are
excluded.  A drop in the helium abundance around $\taus=0.1$ is
related to the dominant role of the $\nu_e$ energy density, since the
spectrum distortion is shifted to smaller energies.

\begin{figure}[hbt]
\psfig{file=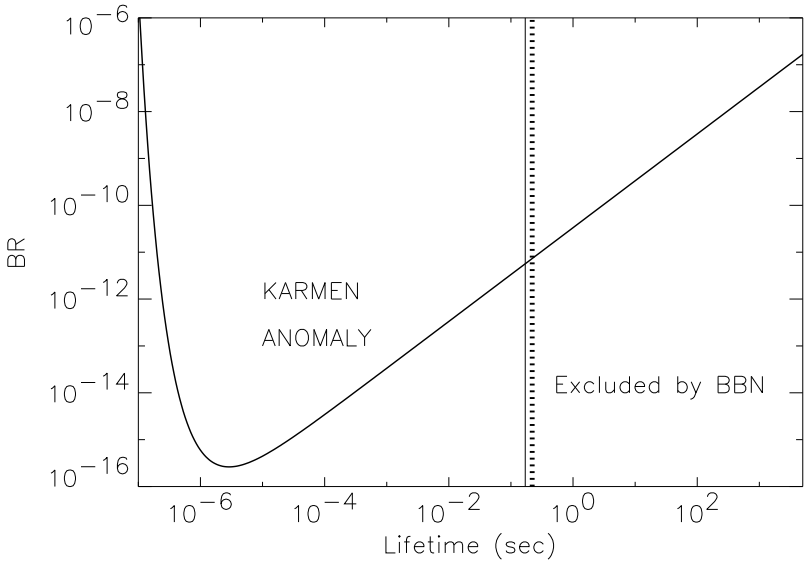,width=4.5in,height=3.2in}
\caption{Branching ratio versus lifetime.  
A comparison of the KARMEN experimental data and the bound obtained here.  
The BBN bound excludes lifetimes bigger than 0.17~sec.
The SN~1987A bound excludes the entire region presented.}
\label{fig:bound}
\end{figure}

Finally in fig.~\ref{fig:bound} we compare the KARMEN experimental 
data with the bound obtained from BBN. One sees that the expected 
new data from the NOMAD
Collaboration together with our BBN bound leave a small allowed window
for a sterile neutrino with a lifetime around 0.1--0.2~sec.

\section{Supernova Limits} 
\subsection{Small Mixing Angle}

The mixing angles of sterile neutrinos with the standard active
flavors is tightly constrained by standard arguments related to
supernova (SN) physics and to the neutrino observations of SN~1987A.
Some of these arguments have been sketched out in the context of the
KARMEN anomaly in Ref.~\cite{barger95}. 

The simplest limit arises from the ``energy-loss argument.''  The
SN~1987A observations imply that a SN core may not emit too much
energy in an ``invisible channel'' as this would unduly shorten the
observed neutrino burst.  Reasonably accurate limits are obtained by
demanding that the ``exotic'' energy-loss rate should
obey~\cite{Raffelt96,Raffelt99}
\begin{equation}\label{eq:snargument}
\epsilon\alt 10^{19}~{\rm erg~g^{-1}~s^{-1}}~,
\end{equation}
where $\epsilon$ is to be calculated at typical average conditions of
a SN core ($\rho=3\times10^{14}~{\rm g~cm^{-3}}$, $T=30~{\rm MeV}$).

Sterile neutrinos $\nus$ are produced because they mix with one of
the standard ones. If the standard neutrino masses are all in the
sub-eV range, the assumed sterile neutrino mass of
$m_{\nu_s}=33.9~{\rm MeV}$ assures a mass difference so large that
medium effects on the oscillations can be neglected. The oscillation
frequency is so large that a standard neutrino $\nu_a$, once produced,
oscillates many times before collisions interrupt the coherent
development of the flavor amplitude. The average probability of
finding the original $\nu_a$ in the $\nu_s$ flavor state is
$\frac{1}{2}\sin^2(2\Theta)$ where $\Theta$ is the
$\nu_a$-$\nu_s$-mixing angle. Oscillations are interrupted with the
collision rate $\Gamma$ of the $\nu_a$ flavor, leading to the standard
sterile-neutrino production rate of
$\frac{1}{2}\sin^2(2\Theta)\Gamma$. 

We estimate $\Gamma$ as the neutral-current collision rate on free
nucleons, ignoring correlation and degeneracy effects, so that:
\begin{equation}
\Gamma=\frac{C_V^2+3C_A^2}{\pi}\,G_F^2 n_B E_\nu^2~,
\end{equation}
with $G_F$ the Fermi constant and $n_B$ the number density of baryons
(nucleons). For a mix of protons and neutrons we use an average
neutral-current coupling constant of $(C_V^2+3C_A^2)\approx 1$.  If we
further assume that the trapped active neutrinos do not have a
significant chemical potential (true for $\nu_\mu$ and $\nu_\tau$, but
not for $\nu_e$), the energy-loss rate is
\begin{equation}
\epsilon_{\nus}=\frac{\sin^2(2\Theta)}{2}\,\frac{G_F^2}{\pi^3 m_N}
\int_0^\infty dE_\nu\,\frac{E_\nu^5}{e^{E_\nu/T}+1}~,
\end{equation}
where $m_N$ is the nucleon mass. For simplicity we use 
Maxwell-Boltzmann statistics for the neutrinos (we ignore
the $+1$ in the denominator under the integral) and find:
\begin{eqnarray}
\epsilon_{\nus}&=&\frac{\sin^2(2\Theta)}{2}\,\frac{120\,G_F^2 T^6}{\pi^3 m_N}~,
\nonumber\\
&=&\sin^2(2\Theta)\,2.8\times10^{26}~{\rm erg~g^{-1}~s^{-1}}\,
T_{30}^{6}\nonumber\\
\end{eqnarray}
where $T_{30}=T/30~{\rm MeV}$. Comparing this result with
(\ref{eq:snargument}) leads to a limit
\begin{equation}\label{eq:mixinglimit}
\sin^2(2\Theta)\alt 3\times10^{-8}\,T_{30}^{-6}~.
\end{equation}
The temperature $T=30~{\rm MeV}$ is at the lower end of what is found
in typical numerical calculations so that this limit is reasonably
conservative.

For the mixing with $\nu_e$ the limit is more restrictive because the
electron neutrinos in a SN core are highly degenerate, leading to a
larger conversion rate and a larger amount of energy liberated per
collision. Several authors found $\sin^2(2\Theta)\alt 10^{-10}$ for
this case~\cite{Kainulainen91,Raffelt93}.

In our calculation of the emission rate we have taken the $\nu_s$ to
be effectively massless. The average energy of trapped standard
neutrinos is about $3T$ which far exceeds $m_{\nu_s}$. The scattering
cross section scales with $E_\nu^2$, favoring the emission of
high-energy neutrinos. The average $\nu_s$ energy emerging from a SN
core is thus found to be about $5T$ so that neglecting $m_{\nu_s}$ is
a good approximation---the $\nu_s$ are highly relativistic.

The KARMEN experiment implies that the lifetime of $\nu_s$ exceeds
about $0.1~{\rm s}$ so that these particles escape from the SN before
decaying. On the other hand, they must decay on the long way from
SN~1987A to us. If the decay products include $e^+e^-$ pairs one will
also get $\gamma$ rays from inner bremsstrahlung in the decay.  The
non-observation of a $\gamma$ ray burst coinciding with SN~1987A then
leads to further limits~\cite{Raffelt96,Raffelt99}.

The decay products likely would include standard neutrinos which would
have shown up in the detectors. However, because of the high energies
of the sterile neutrinos which are representative of the SN core
temperature, such events would have much larger energies than those
expected from thermal neutrino emission at the neutrino
sphere. Therefore, the emission of sterile neutrinos from the core and
their subsequent decay cannot mimic the standard SN neutrino signal.

\subsection{Large Mixing Angle}

These upper limits on the mixing angle are only valid if the sterile
neutrinos actually escape from the SN core after production, a
condition that is not satisfied if one of the mixing angles is too
large. Since the mixing angle between $\nu_e$ or $\nu_\mu$ and $\nu_s$
is already constrained from laboratory experiments to be small in this
sense, we worry here only about the $\nu_\tau$-$\nu_s$ mixing
angle. We continue to assume that the standard-neutrino mass
eigenstates are small so that the mass difference and hence the
oscillation frequency between $\nu_\tau$ and $\nu_s$ will be large
compared to a typical collision rate in a SN core and that the mixing
angle in the medium is identical to the vacuum mixing angle.
Therefore, after production the chance of finding a $\nu_s$ in the
active $\nu_\tau$ state will be given by the average value
$\frac{1}{2}\sin^2(2\Theta)$. If the mixing angle is large this will
mean that the sterile neutrino essentially acts as an active one.
It will be trapped, and the energy-loss argument is not applicable.

If the mixing angle is not quite maximal, sterile neutrinos will still
be trapped, but their mean free path will be larger than that of an
active flavor. Energy is transported out of a SN core by neutrino
diffusion, a mechanism which is more effective if the mfp is larger.
Simply put, a neutrino can transport energy over distances of order
the mfp so that more distant regions are thermally coupled if the mfp
becomes larger. Most of the transporting of energy is done by the most
weakly coupled particles which are still trapped. For example, in a SN
core the photon contribution to the energy transport is negligible
because their mfp is very much smaller than that of standard
neutrinos. Conversely, the contribution of sterile neutrinos will {\it
increase\/} with an increasing mfp, i.e.~with a decreasing
$\sin^2(2\Theta)$. 

The effect on the SN~1987A signal will be identical to the effect of
freely escaping sterile neutrinos, i.e.~the signal will be
shortened. We stress that the signal duration is determined by the
diffusion time scale throughout the star. Therefore, increasing the
mfp in the deep interior of the star shortens the cooling time scale.
In a numerical study~\cite{Keil95} the neutrino opacities were
artificially decreased. It was found that the efficiency of
neutrino transfer in the star should not be more than about twice the
standard value to remain consistent with the SN~1987A signal
characteristics. Likewise, in a different study~\cite{Burrows90}
the number of standard neutrino flavors was artificially increased,
again leading to an increased efficiency of energy transfer.
Doubling the effective number of neutrino flavors appears excluded
from the SN~1987A data.

The interaction rate of sterile neutrinos is that of a standard
$\nu_\tau$, times $\frac{1}{2}\sin^{2}(2\Theta)$.  Conversely, the
collision rate for $\nu_\tau$ is $1-\frac{1}{2}\sin^{2}(2\Theta)$
times the standard rate because a $\nu_\tau$ has an average chance
$1-\frac{1}{2}\sin^{2}(2\Theta)$ of being measured as a $\nu_s$.
Assuming that the standard transfer of energy
is dominated by $\nu_\mu$, $\nu_\tau$ and their anti-particles
(the mfp for $\nu_e$ and $\bar\nu_e$ is much shorter due to 
charged-current reactions), then adding the sterile neutrino will
enhance the rate of energy transfer by a factor
\begin{equation}\label{eq:energyfactor}
\frac{1}{1+1}\left(1+\frac{1}{1-\frac{1}{2}\sin^{2}(2\Theta)}
+\frac{1}{\frac{1}{2}\sin^{2}(2\Theta)}\right)~.
\end{equation}
Maximal mixing corresponds to $\sin^2(2\Theta)=1$, implying that both
$\nu_\tau$ and $\nu_s$ each scatter with half the standard rate, so
their mfp is each increased by a factor of 2. Moreover, the sterile
neutrino contributes a second channel for the transfer of
energy. Therefore, the energy flux carried by maximally mixed $\nu_s$
and $\nu_\tau$ is four times that carried by a standard
$\nu_\tau$. This explains the limiting behavior of
(\ref{eq:energyfactor}) for maximal mixing.

The new experimental limit on the $\nu_s$-$\nu_\tau$ mixing angle is
$\sin^2(2\Theta)<10^{-2}$. Therefore, the efficiency of energy
transfer out of a SN core would be enhanced by more than two orders of
magnitude. Such an enhancement is certainly not compatible with the
SN~1987A signal, implying that the mixing angle has to be very small,
i.e.~that it must obey (\ref{eq:mixinglimit}). Therefore, the
loop hole of a large $\nu_s$-$\nu_\tau$ mixing angle has been plugged
by the new experimental constraints.

\section{Conclusion}

Cosmological and astrophysical arguments seem to exclude the
interpretation of the KARMEN anomaly by an unstable sterile neutrino
mixed with $\nut$.  The arguments based on SN~1987A are stronger than
the nucleosynthesis bound. The supernova limit is given by
(\ref{eq:mixinglimit}), i.e.~$\taus > 6\times 10^4$ sec.  This result,
together with the direct experimental constraints on $\nut$,
completely excludes a 33.9~MeV sterile neutrino.  Primordial
nucleosynthesis permits to exclude roughly $\taus >0.2 $ sec.  So with
the existing direct experimental limits, some window, $\taus =
0.1-0.2$ sec, remains open.  The simplifications that we made in
deriving the BBN bound typically lead to a weaker result, so the real
bound may be somewhat stronger. However, if exact calculations confirm
the decrease of ${}^4$He around $\taus = 0.1$ found in this paper,
then BBN will never exclude lifetimes $\sim 0.1$ sec.

The effect happens to be surprisingly sensitive to usually neglected
phenomena, particularly the shape of the $\nue$ spectrum and to the
non-adiabatic variation of temperature.  We hope to do exact (but
rather long) calculations later. Together with a possible improvement
of the observational data on light element abundances and a better
understanding of the subsequent changes of light elements in the
course of cosmological evolution, the BBN bound may become competitive
with the supernova one. If so, the very attractive hypothesis of a
33.9~MeV sterile $\nus$ would be killed by two independent
arguments. Moreover, one may imagine a case when the supernova
arguments are not applicable, while the nucleosynthesis ones still
operate. For example, if $\nus$ possesses an anomalously strong
interaction with nucleons (stronger than the usual weak one), then it
would not noticeably change the energetics of supernovae that we
described here, but would affect nucleosynthesis practically at the
same level as discussed in section~\ref{ns}.  This new interaction
might be related to the anomalously high mass of $\nus$. If this is
the case then a small window for a 33.9~MeV $\nus$ with lifetime
$\tau_{\nu_s} = 0.1$--0.2~sec, still exists at the present time, and
stronger experimental bounds, as well as more accurate calculations of
the impact of $\nus$ on primordial nucleosynthesis, are needed.

\section*{Acknowledgment} 
We are grateful to N.~Krasnikov who brought this subject to our
attention.  In Munich, this work was partly supported by the
Deut\-sche For\-schungs\-ge\-mein\-schaft under grant No.\ SFB 375.
DS thanks the NOMAD collaboration, in particular L.~DiLella,
L.~Camilleri, S.~Gninenko and A.~Kovzelev, for the invitation and
hospitality during their December-1999 collaboration meeting where
this work was presented.

\begin{table}[ht]
\begin{tabular}{|cccc|c|c|} 
  \hline\hline &&&&&\\
& {\bf Process }&&&{\bf S }& {\bf $2^{-5} G^{-2}_{F} |U_{s\tau}|^{-2} S \left| A \right| ^{2}$}\\
&&&&&\\
 \hline\hline
&&&&&\\
$\nu _{s}  $ & $\rightarrow$ & $ \nu_{\tau} + \nu _{\tau} + \bar{\nu}_{\tau} $&&1/2&
$2 (p_{1} \cdot p_{4}) (p_{2} \cdot p_{3})$ \\
\hline
&&&&&\\
$\nu _{s} $ & $\rightarrow$ &
$ \nu_{\tau} + \nu _{e (\mu)}+\bar{\nu}_{e (\mu)} $&&1&
$ (p_{1} \cdot p_{4}) (p_{2} \cdot p_{3}) $\\
\hline
&&&&&\\
$\nu _{s}  $ & $\rightarrow$ & $  \nu_{\tau} + e^{+} + e^{-}$ &&1&
$4[ ( \tilde{g}_{L}^{2} (p_{1} \cdot p_{4}) (p_{2} \cdot p_{3}) $\\
&&&&&$+ g_{R}^{2}  (p_{1} \cdot p_{3}) (p_{2} \cdot p_{4})$ \\
&&&&&$+ \tilde{g}_{L} g_{R} m _{e}^{2} (p_{1} \cdot p_{2})]$\\
\hline
\hline
\end{tabular}

\vskip0.5cm
{\bf Table 1: } Matrix elements for decay processes;
 $\tilde{g}_{L} =g_{L} - 1 = 
- \frac{1}{2} + \sin^{2}\theta_{W}$ and
$g_{R} = \sin^{2} \theta_{W}$.
\end{table}

\begin{table}[ht]
\begin{tabular}{|cccc|c|c|} 
 \hline\hline
&&&&&\\
& {\bf Process }&&&{\bf S }& {\bf $2^{-5} G^{-2}_{F}|U_{s\tau}|^{-2} S \left| A \right| ^{2}$}\\
&&&&&\\
 \hline\hline
&&&&&\\
$\nu _{s} + \bar{\nu}_{\tau} $ & $\rightarrow$ & $  \nu _{\tau} + \bar{\nu}_{\tau} $&&1&
$4 (p_{1} \cdot p_{4}) (p_{2} \cdot p_{3})$ \\
\hline
&&&&&\\
$\nu _s + \nu_\tau $ & $\rightarrow$ & $  \nu _\tau + \nu_\tau $&&1/2&
$2  (p_{1} \cdot p_{2}) (p_{3} \cdot p_{4}) $\\
\hline
&&&&&\\
$\nu _{s}+\bar{\nu}_{\tau} $ & $\rightarrow$ &
$ \nu _{e (\mu)}+\bar{\nu}_{e (\mu)} $&&1&
$ (p_{1} \cdot p_{4}) (p_{2} \cdot p_{3}) $\\
\hline
&&&&&\\
$\nu_{s} + \bar{\nu}_{e (\mu)} $ & $\rightarrow$&
$\nu _{\tau}+\bar{\nu}_{e (\mu)} $&&1&
$ (p_{1} \cdot p_{4}) (p_{2} \cdot p_{3})$\\
\hline
&&&&&\\
$\nu _{s} + \nu_{e (\mu)} $ & $\rightarrow$ &
$ \nu _{\tau}+\nu_{e (\mu)} $&&1&
$ (p_{1} \cdot p_{2}) (p_{3} \cdot p_{4})$\\
\hline
&&&&&\\
$\nu _{s} + \bar{\nu}_{\tau} $ & $\rightarrow$ & $  e^{+} + e^{-}$ &&1&
$4[ ( \tilde{g}_{L}^{2} (p_{1} \cdot p_{4}) (p_{2} \cdot p_{3}) $\\
&&&&&$+ g_{R}^{2}  (p_{1} \cdot p_{3}) (p_{2} \cdot p_{4})$ \\
&&&&&$+ \tilde{g}_{L} g_{R} m _{e}^{2} (p_{1} \cdot p_{2})]$\\
\hline
&&&&&\\
$\nu _{s} +  e^{-} $ & $\rightarrow$ & $ \nu _{\tau} +  e^{-} $&&1&
$4 [ \tilde{g}_{L}^{2} (p_{1} \cdot p_{2}) (p_{3} \cdot p_{4})$ \\
&&&&&$+g_{R}^{2}  (p_{1} \cdot p_{4}) (p_{2} \cdot p_{3})$ \\ 
&&&&&$-\tilde{g}_{L} g_{R} m _{e}^{2} (p_{1} \cdot p_{3}) ] $\\
\hline
&&&&&\\
$\nu _{s} +  e^{+} $ & $\rightarrow$ & $ \nu _{\tau} +  e^{+} $&&1&
$4 [ g_{R} ^{2} (p_{1} \cdot p_{2}) (p_{3} \cdot p_{4})$ \\ 
&&&&&$+\tilde{g}_{L}^{2} (p_{1} \cdot p_{4}) (p_{2} \cdot p_{3}) $ \\
&&&&&$-\tilde{g}_{L} g_{R} m _{e}^{2} (p_{1} \cdot p_{3}) ] $ \\
\hline
\hline
\end{tabular}
\\
\vskip0.5cm
{\bf Table 2: } Matrix elements for scattering processes;
 $\tilde{g}_{L} =g_{L} - 1 = 
- \frac{1}{2} + \sin^{2}\theta_{W}$ and
$g_{R} = \sin^{2} \theta_{W}$.
\end{table}

\bigskip

\newpage

\indent{\ }

\newpage

\end{document}